\documentclass[aip,jmp,author-year,groupedaddress,nofootinbib,tightenlines]{revtex4-1}

\usepackage{graphicx}
\usepackage{amsmath}

\usepackage[english]{babel}
\usepackage[T1]{fontenc}
\usepackage[ansinew]{inputenc}
\usepackage{lmodern}
\usepackage{units}
\usepackage{dsfont}
\usepackage{amsfonts}
\usepackage{amssymb}
\usepackage{nameref} 
\usepackage[labelfont=bf,figurename=Fig.]{caption}
\usepackage[labelfont=bf,tablename=Table]{caption}
\usepackage[justification=centering]{caption}

\usepackage{color}
\usepackage{url}
\usepackage[nottoc]{tocbibind}
\usepackage{setspace}
\usepackage{bbm}
\usepackage{float}
\usepackage{nomencl}

\begin{document}

\title{Concurrent Credit Portfolio Losses}

\author{Joachim Sicking}
\email{joachim.sicking@uni-duisburg-essen.de}
\affiliation{Faculty of Physics, University of Duisburg-Essen, Lotharstrasse 1, D-47048 Duisburg, Germany} 

\author{Thomas Guhr}
\email{thomas.guhr@uni-duisburg-essen.de}
\affiliation{Faculty of Physics, University of Duisburg-Essen, Lotharstrasse 1, D-47048 Duisburg, Germany} 

\author{Rudi Sch{\"a}fer}
\email{rudi.schaefer@uni-duisburg-essen.de}
\affiliation{Faculty of Physics, University of Duisburg-Essen, Lotharstrasse 1, D-47048 Duisburg, Germany}

\begin{abstract}
We consider the problem of concurrent portfolio losses in two non-overlapping credit portfolios. In order to explore the full statistical dependence structure of such portfolio losses, we estimate their empirical pairwise copulas. Instead of a Gaussian dependence, we typically find a strong asymmetry in the copulas. Concurrent large portfolio losses are much more likely than small ones. 
Studying the dependences of these losses as a function of portfolio size, we moreover reveal that not only large portfolios of thousands of contracts, but also medium-sized and small ones with only a few dozens of contracts exhibit notable portfolio loss correlations. Anticipated idiosyncratic effects turn out to be negligible.
These are troublesome insights not only for investors in structured fixed-income products, but particularly for the stability of the financial sector. 
\end{abstract}

\maketitle

\section{Introduction}
Credit portfolios play a crucial role in the economy and numerous approaches and ideas have been put forward to model the losses that might occur to the portfolio holder, see eg \cite{crouhy}, \cite{egloff}, \cite{bielecki}, \cite{bluhm}, \cite{lando}, \cite{mcneil}. While many contributions have been made regarding single credit portfolios,  little attention has been given to another highly important dimension of credit risk: concurrent losses of different portfolios. Considering the portfolios of major participants in the financial system, concurrent extreme losses might impact their solvencies and thus pose high systemic risks. In addition to this macro-economic aspect, the topic of concurrent credit portfolio losses is equally interesting from an investor's point of view. Buying collateralized default obligations (CDO) allows to hold a ``slice'' of each contract within a given portfolio, see \cite{duffieCDO}, \cite{longstaffCDO}, \cite{BenmelechCDO}. Any investor who is invested in CDOs would thus be severely affected by significant concurrent credit portfolio losses as many of their CDO investments would simultaneously fail making their scheduled payments.    
This raises the question of how strongly the losses of different portfolios are coupled and how their interdependences could be described.
Multiple approaches have been developed to model credit risk. Most of them can be categorized into reduced (see eg \cite{duffieANDsingleton}, \cite{chava}, \cite{schonbucher2003credit}) and structural approaches (see \cite{Merton74}, \cite{elizalde}), which are comprehensively reviewed in \cite{giesecke}. Here, we stick to Merton's structural model that accounts for the empirically observed (\cite{Becker}) dependence of probability of default (PD) and recovery rate (RR) and exhibits a convincing conceptual simplicity.   

In the framework of the Merton model (\cite{Merton74}), the evolution of a credit portfolio can be traced back to multivariate stock price return distributions. This allows us to easily incorporate important features like fat tails and non-stationary asset correlations into credit risk, see eg \cite{schmitt}.
Applying these concepts, we simulate losses of two credit portfolios which are composed of statistically dependent credit contracts. Recalling that correlation coefficients only give full information in the case of Gaussian distributions, we study the statistical dependence of these portfolio losses by means of copulas (\cite{nelsen}). 

The application of copulas in credit risk has thrived in recent years. One of the first and most influential contributions to this field is due to \cite{li}. Li put forward a method to determine default correlations and -- without going into detail -- modeled dependences in credit risk by means of Gaussian copulas. In contrast, we run Monte-Carlo simulations of credit portfolio losses with empirical input and analyze the resulting empirical copulas in detail. Moreover, we study the deviations from Gaussian copulas. Even though Gaussian copulas -- which boil down to a linear correlation of Gaussian variables -- are often inadequate, we will also consider the loss correlation coefficients as a crude measure for the dependence from time to time.
 
In section \ref{sec:model}, we introduce the Merton model of credit risk and relevant statistical concepts. Next, we analyze copulas of simple homogeneous credit portfolios in section \ref{sec:homogeneous} in order to shed light on the mechanisms causing deviations from the often assumed Gaussian dependences. Further, we make the reader aware of some difficulties going along with portfolio loss correlations. Section \ref{sec:empirical} is dedicated to empirical S\&P 500- and Nikkei 225-based credit portfolios and their dependences. We conclude with a discussion of our findings in \mbox{section \ref{sec:discussion}}.

\section{Model and Methods}\label{sec:model}
To make our contribution more widely accessible, we start with an introduction of Merton's model of credit risk and the powerful statistical concept of copulas. Finally, we explain the basics of our numerical simulation.
\subsection{Credit Risk}\label{subsec:creditrisk}
Following \cite{Merton74}, we assume a company's asset value $V_i(t)$ to be the sum of time-independent liabilities $F_i$ and equity $g_i(t)$, $V_i(t) = F_i + g_i(t)$, and model its dynamics by
\begin{equation}\label{eq:stochdiffeq}
  dV_i = \mu_i\,V_i\,dt + \sigma_i\,V_i\,dW_i
\end{equation}
with $\mu_i$ and $\sigma_i$ denoting drift and volatility and $dW_i$ being a random process to be specified. Further, we assume publicly listed companies and thus trace back changes in asset value to stock price returns. This allows us to determine the parameters $\mu_i$ and $\sigma_i$ from stock return time series. 

This deduction has a remarkable weakness though: stock returns do not exclusively reflect the actual changes in the company's equity at any time, as could be assumed. Instead, they are subject to speculation and hence might be an inadequate approximation. However, this drawback is especially important on short return time scales $\Delta t$ -- days or weeks, say -- and is becoming less and less substantial as $\Delta t$ approaches longer time scales. Hence, calculations on annual horizons which are considered hereafter are in line with the reasoning set out above.

Credit risk can be easily introduced in this framework: let us consider a company with $V_{0i} > F_i$ and liabilities, $F_i$, that mature after one year (without any coupon payments in between). If $V_i(T) > F_i$ holds, the company is able to make the required payments and thus fulfills its obligations. Contrarily, if $V_i(T) < F_i$ holds, the company fails to make the payments, ie it defaults. The normalized loss $l_i$ the creditors are suffering can be expressed as\vspace*{0.3cm}
\begin{equation}
 \vspace*{0.2cm} l_i = \frac{F_i - V_i(T)}{F_i} \:\Theta(F_i-V_i(T)),
\end{equation}
with $\Theta(x)$ denoting the Heaviside step function.

In the following, all loss simulations refer to a one-year time horizon and are carried out using the time parameter $T = 252\:\text{days} = 12\times21$ days. This is due to the described correspondence between asset value and stocks and the fact that a calender year has about $252$ trading days, ie the stock markets are open $252$ days a year. Moreover, it is useful to introduce the ratio of liabilities and asset value, $F_i/V_i(t)$, the so-called \textit{leverage}. The default criterion then reads $F_i/V_i(T) < 1$ (no default) and $F_i/V_i(T) > 1$ (default), respectively.   

Next, we pass on from single credit contracts to credit portfolios. The importance of a single contract in a portfolio is determined by comparing the face value of this specific contract with the sum of all face values. To this end we use the fractional face values
\begin{equation}
 f_i = \frac{F_i}{\sum_{j = 1}^{K}{F_j}}\,.
\end{equation}
Then we can write the \textit{normalized portfolio loss} $L$ as
\begin{equation}\label{eq:pfloss}
 L = \sum_{i=1}^{K}{f_i\,l_i} \ \ \ \text{with}\ \ \ l_i = \frac{F_i - V_i(T)}{F_i}\:\Theta(F_i-V_i(T))\,.
\end{equation}
Here, $i$ enumerates the $K$ credit contracts. $K$ is hereafter often referred to as \textit{portfolio size}. Due to normalization, $L$ ranges from zero to one. Further note that all credit contracts are supposed to mature at time $T$ and are to be coupon-free, ie the total face value must be payed at maturity. 

According to the explanations above, the dynamics of such a portfolio of correlated credit contracts is well described by correlated stock price returns, which are frequently modeled using a multivariate normal distribution
\begin{equation}\label{eq:multinormal}
g(r|\Sigma) = \frac{1}{\sqrt{\det(2\pi\Sigma)}}\exp\left(-\frac{1}{2} r'\,\Sigma^{-1} r\right).
\end{equation}
Here, $r = (r_1,r_2,...,r_K)$ denotes the $K$-dimensional return vector and $r'$ its transpose.
The components of $r$ are coupled via a $K\times K$ covariance matrix $\Sigma$, which can be expressed as $\Sigma = \sigma\,C\,\sigma$ with $C$ being the $K\times K$ correlation matrix and $\sigma = \text{diag}(\sigma_1,\sigma_2,...,\sigma_K)$ a diagonal matrix containing the volatilities $\sigma_i$ of the different stock returns $r_i$. Assuming returns according to Eq.~(\ref{eq:multinormal}) results in a multivariate log-normal distribution for the stock prices. 

More realistic, however, is the aforementioned random matrix concept of \cite{schmitt}, an extension of \cite{muennix2}, which accounts for non-stationary covariances. Here, Eq.~(\ref{eq:multinormal}) with the stationary covariance matrix $\Sigma$ serves as a starting point. Averaging over covariance matrices fluctuating around $\Sigma$ in the course of time finally yields the multivariate ensemble averaged distribution
\begin{equation}\label{eq:ensembledist}
 \langle g\rangle(r|\Sigma,N) = \int\limits_{0}^{\infty}{dz\:\chi_N^2(z)\:g\left(r\Big|\frac{z}{N}\Sigma\right)}.
\end{equation}
Here, the parameter $N$ controls the strength of fluctuations around the mean $\Sigma$. Compared to the multivariate normal distribution with a fixed covariance matrix, the ensemble averaged distribution exhibits fat tails for any finite $N$, the more pronounced the smaller $N$. $N$ is determined by fitting and depends strongly on the considered return interval $\Delta t$. \cite{schmitt} have shown particularly that daily returns are matched best by $N = 5$, while annual returns behave normally ($N \gg 1$).
Yet, the ensemble approach remains useful if simplified correlation matrices with homogeneous off-diagonal elements ${C_{ij} = \mathrm{Corr}(V_i, V_j) = c_\mathrm{a}}$, $i\,\neq j$, are used\footnote{The subscript “a” stands for “asset” and is used to distinguish
asset correlations $c_\mathrm{a}$ from other correlation coefficients.}. \cite{schmitt3} have shown that distributions of annual stock returns are fitted best by $N = 5$ in this case, as opposed to $N \gg 1$ in the case of full correlation matrices.

According to It\={o}'s lemma (\cite{ito}), we obtain the asset values at maturity $V_i(T)$ from the stock price returns $r_i$ via
\begin{equation}\label{eq:V_T}
 V_i(T) = V_{0i}\:\exp\left[r_i + \left(\mu_i - \frac{\sigma_i^2}{2}\right)T\right].
\end{equation}
\subsection{Copulas}\label{subsec:copulas}
Let us assume a bivariate random variable $X = (X_1,X_2)$ that is comprehensively described by means of its joint probability density, $f(x_1,x_2)$, say. The main idea of copulas, which were introduced by \cite{sklar59}, is to separate the statistical dependence of $X_1$ and $X_2$ from the two marginal distributions. This is achieved by the copula
\begin{equation}
 \text{Cop}(u,v) = F(F_1^{-1}(u),F_2^{-1}(v)),
\end{equation}
which is the cumulative joint distribution of the quantiles $u$ and $v$ with $(u,v)\in [0,1]\times [0,1]$. $F_i^{-1}$ denotes the inverse marginal cdfs.  This construction -- the copula as a composition of inverse marginal cdfs and the joint cdf -- allows us to analyze statistical dependences regardless of the marginal distributions.
Instead of the copula itself, we will work with the \textit{copula density} $\text{cop}(u,v)$, which is given by
\begin{equation}
 \text{cop}(u,v) = \frac{\partial^2 \text{Cop}(u,v)}{\partial u \partial v}.
\end{equation}
In particular, we are interested in estimating and analyzing copula densities obtained from empirical or simulation data. The construction of such an empirical copula density from a bivariate data set of length $n$ takes the following steps. First, we need to replace the actual values $(x_i,y_i), i \in \{1,...,n\}$ with their ranks $(\text{rank}(x_i),\text{rank}(y_i))$; this is conducted separately for $x_i$- and $y_i$-values. Next, the ranked values, $1\leq \text{rank}(x_i)\leq n$ are scaled to the interval $[0,1]$. These normalized ranks produce the empirical cdf-values of the initial data points in the data set.

Binning these new data points yields a two-dimensional histogram which is -- if normalized appropriately -- an empirical approximation of the copula density. The number of bins is chosen to be $b = 20$ in each direction.
\subsection{Simulation Setup}\label{subsec:setup}
\begin{figure}[h]
 \centering
 \includegraphics[width=0.6\textwidth]{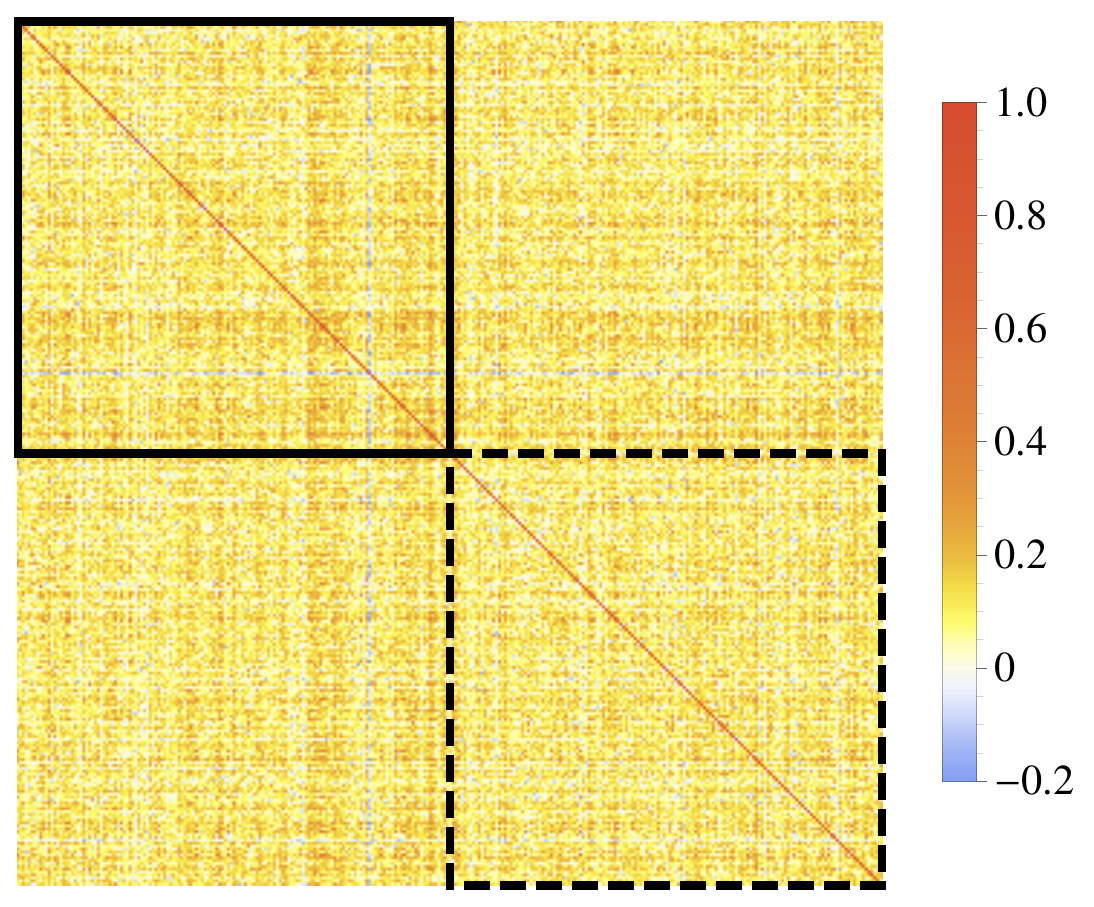}
  \caption{Exemplary correlation matrix illustrating a financial market. The black-rimmed squares (solid and dashed) correspond to two disjoint portfolios.}
 \label{fig:Two_Portfolios_Schematic}
\end{figure}
Here and in all simulations hereafter, we consider two credit portfolios which are set up according to Fig. \ref{fig:Two_Portfolios_Schematic}. On a given financial market, illustrated by means of its correlation matrix, we choose these portfolios to be non-overlapping, ie no credit contract is part of both portfolios. The number of contracts in each portfolio is $K$, ie they are of equal size. We mark the two portfolios in Fig. \ref{fig:Two_Portfolios_Schematic} as black-rimmed squares. We note that the off-diagonal squares are important as well. They illustrate the asset correlations of companies which are part of different portfolios and thus cause interdependences between the two portfolio losses. 

The portfolio loss copula of these portfolios is obtained as follows: We simulate the correlated asset values, $V_i(T)$ with $T = 252$, of all companies within the financial market. For both portfolios, we choose the affiliated $V_i(T)$ and calculate the corresponding portfolio losses $L_1$ and $L_2$ according to Eq.~(\ref{eq:pfloss}). Repeating this portfolio loss calculation a few thousand times yields enough data to estimate the copula histogram of the two portfolio losses. 

We repeat these steps several thousand times for different pairs of portfolios in order to avoid results which are due to the specific features of two particular portfolios. All these pairs are non-overlapping and operate on the same financial market; each portfolio is of size $K$. We determine the copula histogram in each case and average over them. The resulting mean copula histogram provides information about the average portfolio interdependence. However, it still depends on the portfolio size $K$ as well as on the considered market.

As mentioned above, statistical dependences are often reduced to correlation coefficients or -- at the level of copulas -- to Gaussian copulas. Here, we would like to study the deviations of the averaged empirical copula from the related Gaussian copula. In order to achieve this Gaussian copula, we need to estimate its parameter $c$ in the first place. Here, $c$ equals the averaged correlation coefficient of portfolio losses $C_{L_1L_2} = \mathrm{Corr}(L_1,L_2)$, which is determined -- similarly to the averaged copula histogram -- by averaging over the portfolio loss correlation for every portfolio pair.

Moreover, we employ leverages, which are drawn from a uniform distribution $\mathcal{U}$ on the interval $[0.6,0.9]$,
\begin{equation}\label{eq:levdist}
  \frac{F_i}{V_{0i}} \sim \mathcal{U}(0.6,0.9),\ \ \ i = 1, ..., K,
\end{equation}
for every non-homogeneous, ie heterogeneous, portfolio hereafter.

\section{Simulation of Homogeneous Credit Portfolios}\label{sec:homogeneous}
We shall first restrict ourselves to the simulation of homogeneous credit portfolios in order to systematically study the impact of different parameters on portfolio loss copulas. In particular, we focus on asset correlations and drifts. In addition to copulas, we analyze the wide-spread measure of portfolio loss correlations. Here, our focus is the dependence of portfolio loss correlations on the underlying asset correlations and on the portfolio size. 

\subsection{Impact of Asset Correlations on Portfolio Loss Copulas}\label{subsec:assetcorr}
\begin{figure}[h]
 \centering
 \includegraphics[width=0.65\textwidth]{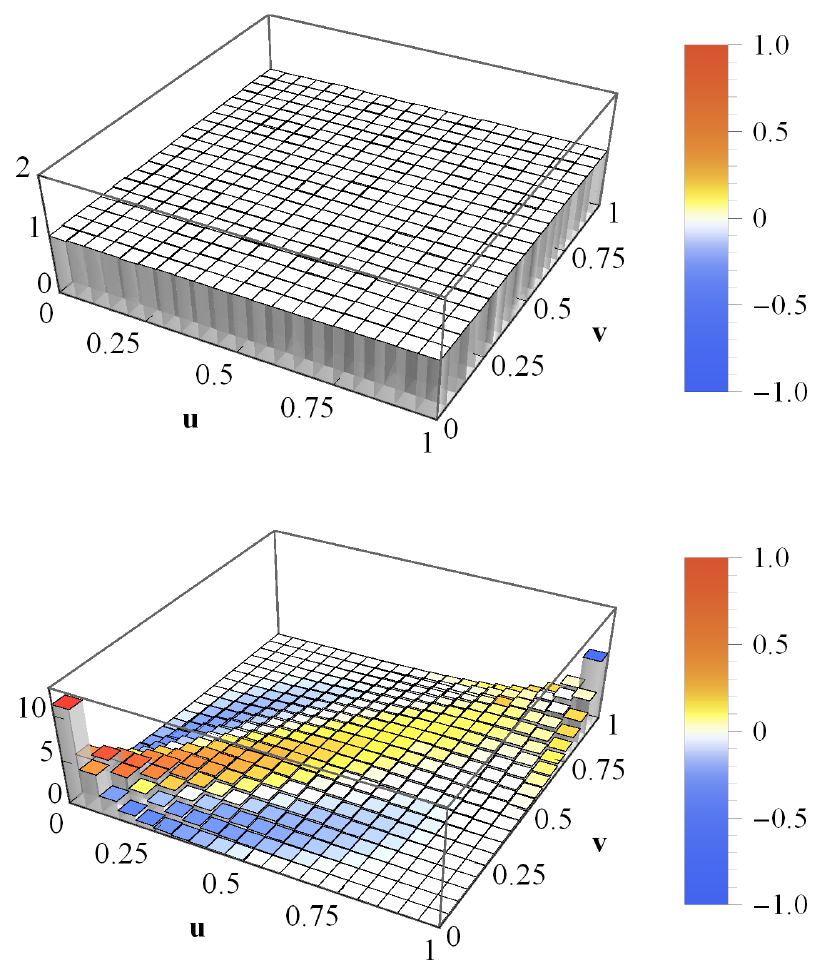}
  \caption{Averaged copula histograms of portfolio losses $L_1$ and $L_2$ of homogeneous portfolios with vanishing (mean) asset correlations, ie $c_\mathrm{a} = 0$. The asset values are multivariate normal with $N \rightarrow \infty$ (top) and multivariate heavy-tailed with $N = 5$ (bottom). The coloring indicates the local deviations from the Gaussian copula with ${c = C_{L_1L_2} = 0.752}$.}
 \label{fig:Copulas_Homogeneous_c=0_Differentns_cropped}
\end{figure}

For the simulation of asset value processes and loss calculation, we specify $\mu = 10^{-3}\:\text{day}^{-1}$, $\sigma = 0.03\:\text{day}^{-1/2}$ and leverages $F/V_0 = 0.75$. Our choice of rather small portfolios, $K = 50$, will be discussed at the end of this section. Furthermore, we assume a market with vanishing asset correlation, $c_\mathrm{a} = 0$.
The simulation of asset values $V_i(T)$, which underlies the portfolio loss calculation and thus the averaged copula histogram, is run in two different ways. On the one hand, we assume Gaussian dynamics, ie we use -- in the framework of the ensemble averaged distribution -- the ``fat-tail'' parameter $N \rightarrow \infty$. On the other hand, we choose $N = 5$ -- in accordance with the findings of \cite{schmitt3} for homogeneous average correlation matrices and annual time horizons. The resulting averaged copula histograms for $10\,000$ portfolio loss simulations and $1000$ portfolio pairs\footnote{The choice of different portfolio pairs is unnecessary for homogeneous portfolios, as they are all equivalent. However, it matters for heterogeneous portfolios and we employ -- except for the input parameters -- the same simulation in either case.} are shown in Fig. \ref{fig:Copulas_Homogeneous_c=0_Differentns_cropped}. 

For $N \rightarrow \infty$ (top panel), we find an independence copula, which exhibits a constant level of one due to normalization. We observe only minimal deviations due to the finite simulation length.
Measuring the average portfolio loss correlation yields $C_{L_1L_2} = 0$ and thus the related Gaussian copula is an independence copula as well. The reason for this result is quite obvious: the simulation of asset values is based on uncorrelated Gaussian random numbers, ie they are statistically independent. Unsurprisingly, quantities derived from these random numbers do not exhibit dependences either.

The combination of $c_\mathrm{a} = 0$ and $N = 5$ yields a different result (bottom panel of Fig. \ref{fig:Copulas_Homogeneous_c=0_Differentns_cropped}); its deviations from the independence copula are striking. Ignoring the coloring for a moment and considering solely the bar heights is sufficient to recognize that this copula is not Gaussian: the bar in the $(0,0)$-corner is twice as high as its equivalent in the $(1,1)$-corner. Contrarily, Gaussian copulas are perfectly symmetric regarding the line spanned by $(0,1)$ and $(1,0)$. Nevertheless, we calculate the correlation coefficient $c = C_{L_1L_2} = 0.752$ of the simulated portfolio losses and determine the corresponding Gaussian copula.  It is not shown on its own, but the difference between the actual copula and the Gaussian copula within each bin is illustrated by means of coloring. The color bar (on the right in Fig. \ref{fig:Copulas_Homogeneous_c=0_Differentns_cropped}) provides the translation between color and value. In general, the colors yellow, orange and red indicate that the actual copula exhibits a stronger dependence within the given $(u,v)$-interval than predicted by a Gaussian copula. Contrarily, turquoise and blue are illustrations of local dependences weaker than Gaussian. 

The differences from the independence copula ($N \rightarrow \infty$) are due to the fact that finite values of $N$, eg $N = 5$, cause fluctuating correlations (cf section \ref{sec:model}). Here, they are centered at $c_\mathrm{a} = 0$, ie positive and negative fluctuations are equally likely. This raises the question why the portfolio losses exhibit such a strong positive correlation,  $c = C_{L_1L_2} = 0.752$. The answer to this question is twofold: First, a large correlation matrix with strong positive and negative correlations implies, roughly speaking, a devision of the companies into two blocks with positive correlations within the blocks and negative correlations between them. Second, credit risk is -- as discussed before -- highly asymmetric. There is no positive impact of prospering and thus non-defaulting companies on portfolio losses. The Heaviside functions in Eq.~(\ref{eq:pfloss}) cut off all these non-defaults and project them onto zero. This way, anti-correlations of asset values contribute to the portfolio loss correlation only in a rather limited fashion. Even worse, due to the aforementioned block-structure, the negative correlation between the blocks makes it more likely that one of the blocks is adversely affected. And the positive correlations within the blocks imply a high risk of concurrent defaults.

\subsection{Drift Dependence of Portfolio Loss Copulas}\label{subsec:drift}
\begin{figure}[H]
 \centering
 \includegraphics[width=0.65\textwidth]{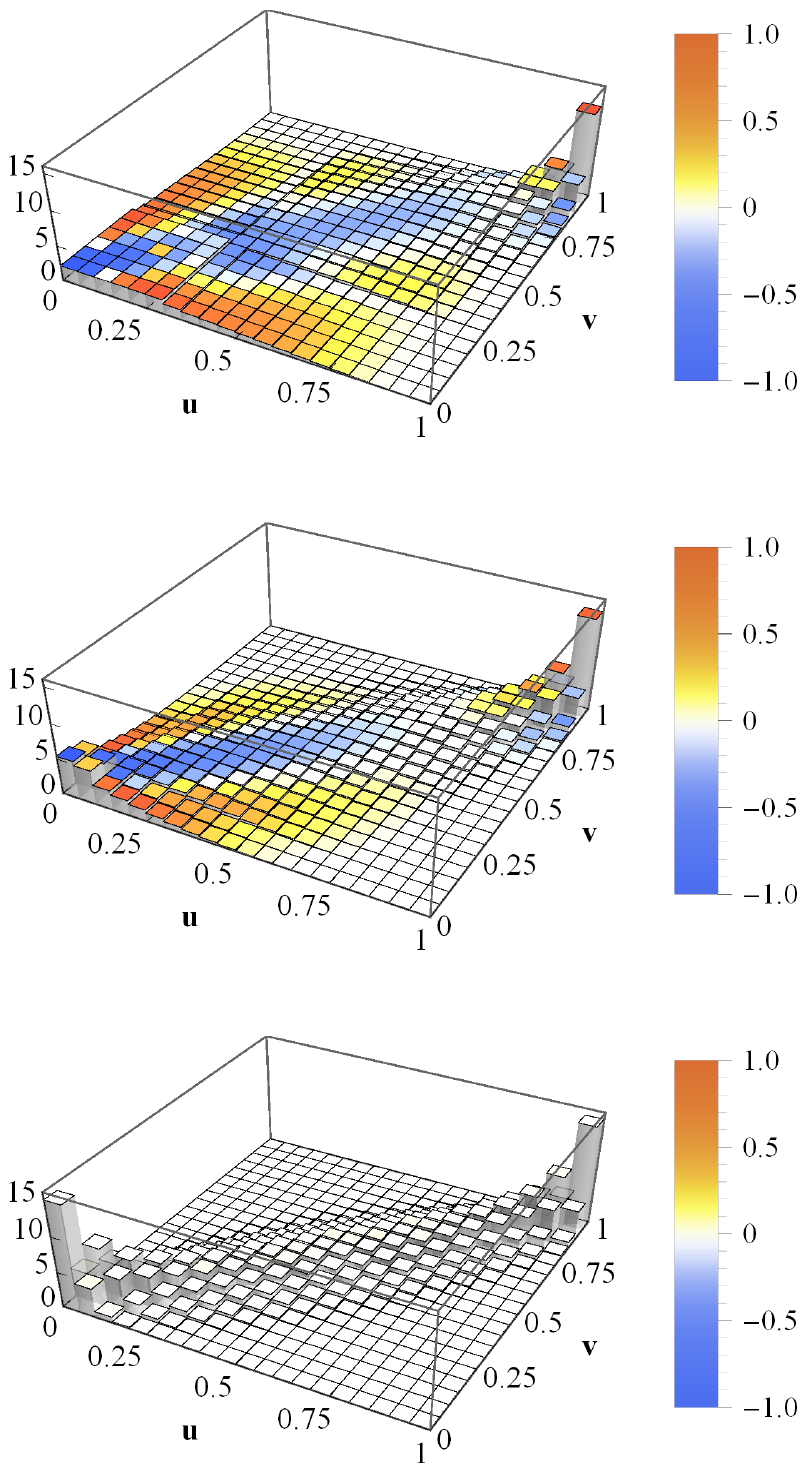}
  \caption{Averaged copula histograms of portfolio losses $L_1$ and $L_2$ of homogeneous portfolios with asset correlations $c_\mathrm{a} = 0.3$. The drifts are ${\mu = 10^{-3}\:\text{day}^{-1}}$ (top), $3\times10^{-4}\:\text{day}^{-1}$ (middle) and ${-3\times10^{-3}\:\text{day}^{-1}}$ (bottom). The coloring refers to the local deviations from the Gaussian copula with $C_{L_1L_2}$. The asset values are multivariate normal ($N \rightarrow \infty$).}
 \label{fig:Copulas_Homogeneous_DifferentMus_cropped}
\end{figure}

Defaulting portfolios ($L > 0$) contribute to the portfolio loss copula as well as non-defaulting portfolios ($L = 0$). We would like to study the impact of each on the portfolio loss copula in greater detail. First, we vary the asset value drifts $\mu$ in order to influence the default-non-default ratio. Deviating from the specifications above, $c_\mathrm{a} = 0.3$, instead of $c_\mathrm{a} = 0$, and $N \rightarrow \infty$ are used. Moreover, we choose the slightly lower volatility $\sigma = 0.02\:\text{day}^{-1/2}$. The resulting averaged copulas for three different drift parameters are shown in Fig. \ref{fig:Copulas_Homogeneous_DifferentMus_cropped}.
\newpage
For $\mu = 10^{-3}\:\text{day}^{-1}$ (top panel), non-defaults occur with a probability of $39.1\%$. The ``plateau'' within $[0,0.3]\times [0,0.3]$ is due to simultaneous non-defaults of both portfolios. More striking, however, is the strong tail dependence in the (1,1)-corner. Even though we observe such an obviously non-Gaussian averaged copula, we estimate the average portfolio loss correlation and obtain $C_{L_1L_2} = 0.851$.

Choosing the smaller drift $\mu = 3\times10^{-4}\:\text{day}^{-1}$ (middle panel) yields -- unsurprisingly -- a lower probability of non-defaults, namely $12.8\%$, and a copula which is far more Gaussian compared to the previous one. Nevertheless, there are several deviations from an ideal Gaussian copula. The $(1,1)$-tail is narrower and more pointed. And even though we have not changed the dependences of asset values, the resulting average portfolio loss correlation is $C_{L_1L_2} = 0.904$ and thus approximately $5\%$ higher compared to the result in the top panel.

Finally, the drift $\mu = -3\times10^{-3}\:\text{day}^{-1}$ leaves no non-default events at all. We observe an ideal Gaussian copula (no coloring except for white) with an average portfolio loss correlation of $C_{L_1L_2} = 0.954$. Again, this portfolio loss correlation is $5\%$ higher compared to the previous result.
\begin{figure}[h]
 \centering
 \includegraphics[width=0.75\textwidth]{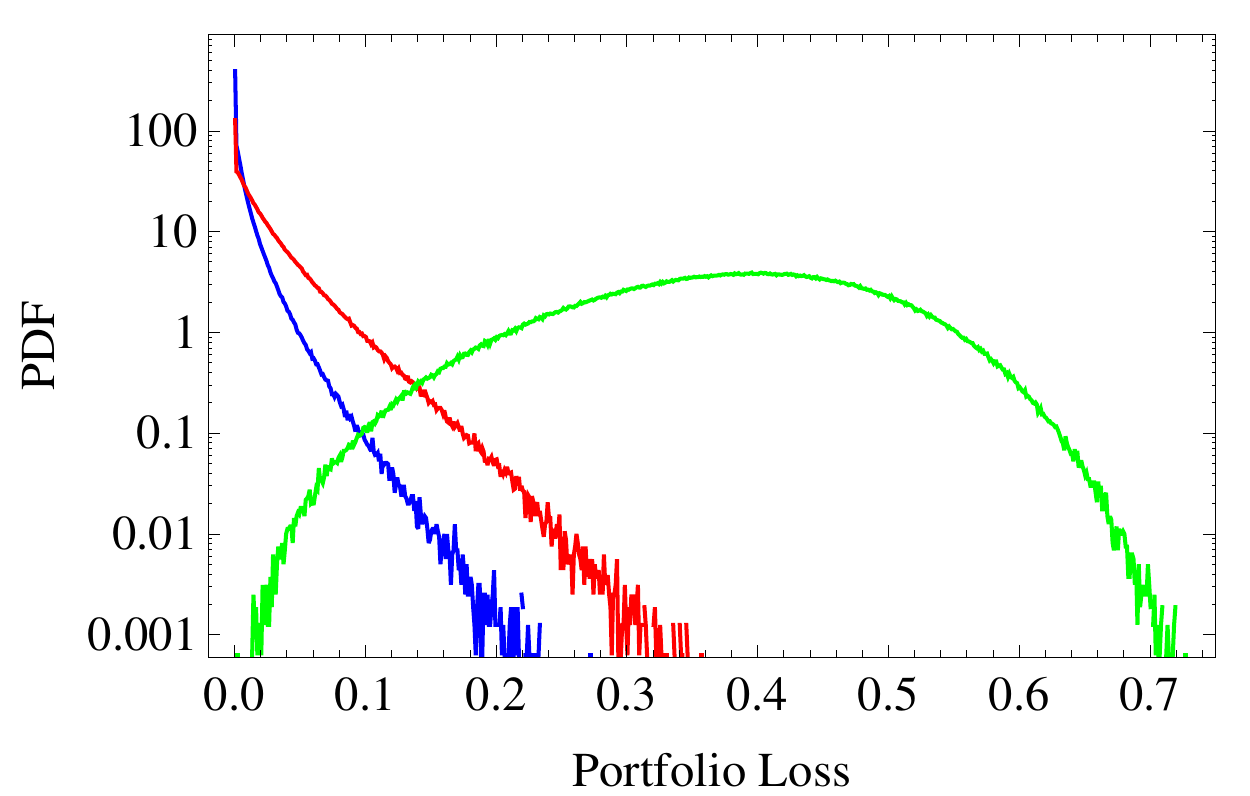}
  \caption{Semi-log scaled portfolio loss pdfs for homogeneous portfolios (asset correlations $c_\mathrm{a} = 0.3$) with drift parameters $\mu = 10^{-3}\:\text{day}^{-1}$ (blue), $3\times10^{-4}\:\text{day}^{-1}$ (red) and $-3\times10^{-3}\:\text{day}^{-1}$ (green). The asset values are multivariate normal ($N \rightarrow \infty$).}
 \label{fig:LossDistributions_Homogeneous_DifferentMus}
\end{figure}
We have seen that portfolio loss correlations increase and that the averaged copula turns ever more Gaussian if the percentage of default events increases. In order to explain these findings, let us analyze the according marginal distributions of the portfolio losses (Fig. \ref{fig:LossDistributions_Homogeneous_DifferentMus}). For $\mu = 10^{-3}\:\text{day}^{-1}$ and $\mu = 3\times10^{-4}\:\text{day}^{-1}$, ie for portfolio default probabilities of $60.9\%$ and $87.2\%$, respectively, we find extremely asymmetric portfolio loss pdfs. Both exhibit a fat tail for high portfolio losses and a delta peak at zero due to non-defaults. In contrast, for $\mu = -3\times10^{-3}\:\text{day}^{-1}$, ie for a default probability of $100\%$, the pdf has no delta peak at zero -- as the chance of non-defaults is $0\%$ -- and moreover is almost symmetric. Nevertheless, it shows some deviations from a normal distribution, even though we observe an ideally Gaussian dependence of the portfolio losses (see bottom panel of Fig.~\ref{fig:Copulas_Homogeneous_DifferentMus_cropped}).

Moreover, let us consider enlarged versions ($K \gg 50$) of these three portfolios for a moment. It can be easily shown that such scalings lead to a decreasing number of non-default events and thus to smaller delta peaks of the portfolio loss pdfs at zero, which vanish entirely if $K$ is large enough. Despite this change, however, the shapes of the pdfs remain almost the same for $K \gg 50$. In contrast to this virtual $K$-independence of the marginal distributions, the statistical dependences turn ever more Gaussian as $K$ increases and the number of non-default events decreases.

Taking all these findings into consideration, we infer that the loss of information -- resulting from the projections of non-default events onto zero -- is responsible for the decrease in portfolio loss correlation with decreasing likeliness of default events. Furthermore and even more importantly, we deduce that these projections onto zero cause the observed deviations of the statistical dependences from Gaussian copulas.

\subsection{Portfolio Loss Correlation}\label{subsec:losscorr}
Even though non-Gaussian portfolio loss copulas cannot be described comprehensively by means of correlation coefficients, correlations are frequently applied in credit risk, eg by \cite{lucas} and \cite{li}. It is worth pointing out that our concept of loss or default correlations differs from theirs, as we account for the actual loss values. In contrast, Lucas applies the frequently used ``binary'' approach, ie he only distinguishes default and non-default, while Li's concept is based on so-called ``survival times''.

\begin{figure}[h]
 \centering
 \vspace*{1cm}
 \includegraphics[width=1.0\textwidth]{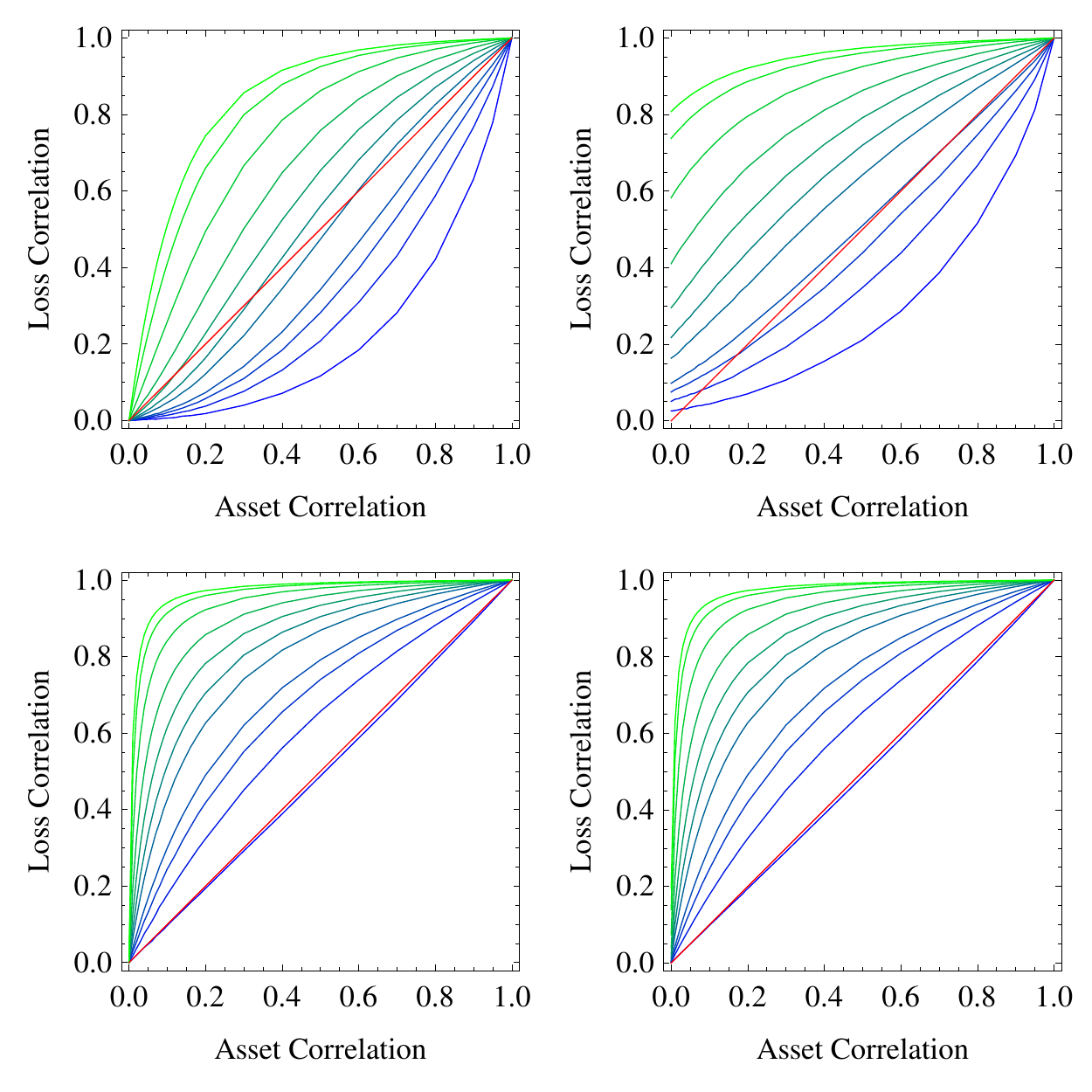}
  \caption{Portfolio loss correlation $C_{L_1L_2}$ as a function of asset correlation $c_a$. The size $K$ of the underlying homogeneous portfolio ranges from $1$ (blue) over $2,3,4,7,10,15,25,50,100$ to $150$ (green). The daily drifts are set to $\mu = 2\times10^{-3}\:\text{day}^{-1}$ (top row) and $\mu = -3\times10^{-3}\:\text{day}^{-1}$ (bottom row). All asset values are multivariate normal with $N \rightarrow \infty$ (left column) and multivariate heavy-tailed with $N = 5$ (right column). The bisecting line is shown in red.}
 \label{fig:LossCorr_AssetCorr_K_Homogeneous}
\end{figure}

Here, we examine portfolio loss correlations more closely, because the reduction of an entire statistical dependence to a single number simplifies the analyses of further parameter dependences considerably. Especially, we are interested in the dependence of portfolio loss correlations on the underlying asset correlations and on the portfolio size $K$. They are shown in Fig. \ref{fig:LossCorr_AssetCorr_K_Homogeneous} for a homogeneous portfolio -- as described above -- with drifts $\mu = 2\times10^{-3}\:\text{day}^{-1}$ (top row) and $\mu = -3\times10^{-3}\:\text{day}^{-1}$ (bottom row) and ``fat-tail'' parameters $N \rightarrow \infty$ (left column) and $N = 5$ (right column). Portfolio size $K$ serves as a curve parameter and takes on values between $K = 1$ and\footnote{For a list of all portfolio sizes $K$, cf caption of Fig. \ref{fig:LossCorr_AssetCorr_K_Homogeneous}.} $K = 150$. 

First, we find the portfolio loss correlation to be a monotonic function of asset correlation as well as of portfolio size -- regardless of drift $\mu$ and ``fat-tail'' parameter $N$. Keeping in mind the results of the previous subsection, we already know that correlation coefficients capture the entire dependence of portfolio losses for $\mu = -3\times10^{-3}\:\text{day}^{-1}$ and $N \rightarrow \infty$ (bottom left panel). Obviously, asset and loss correlation equal each other for $K = 1$. As $K$ increases, portfolio loss correlation is a concave function of asset correlation and increases much more steeply. An asset correlation differing only slightly from zero is sufficient to make it take on values close to one. This behavior is related to credit portfolio granularity, which increases with portfolio size $K$ and causes ever more deterministic results. In the limit $K \rightarrow \infty$ and for every $c_\mathrm{a} > 0$, all portfolio losses are the same and thus are perfectly correlated. In order to preserve some randomness to achieve non-trivial results, we decided to choose the rather small portfolio size of $K = 50$ in the previous subsections. 

For $\mu = 2\times10^{-3}\:\text{day}^{-1}$ and $N \rightarrow \infty$ (top left panel of Fig. \ref{fig:LossCorr_AssetCorr_K_Homogeneous}), portfolio loss correlations are only a first approximation of the actual statistical dependence of portfolio losses. 
\clearpage
Here, concavity holds only for large $K$ (namely, $K = 50,100,150$), while portfolio loss correlation is a convex function of asset correlation for small $K$ (namely, $K = 1,2,3,4$). The functional dependences for all $K$ in between are neither concave nor convex. If convexity holds, we find the portfolio loss correlations to be always smaller than the corresponding asset correlations.

Comparing the results for $N \rightarrow \infty$ in the top and bottom left panel reveals that the strength of portfolio loss correlation is not only a function of asset correlation, but moreover a function of the ratio of non-default and default events. This result is of particular importance for stress testing, as it implies that the correlations of single credit losses as well as of credit portfolio losses are stronger during financial crises than in calm market periods!

Contrasting these results with the results for $N = 5$ (right column), we find remarkable deviations for $\mu = 2\times10^{-3}\:\text{day}^{-1}$. We observe large portfolio loss correlations for $c_\mathrm{a} = 0$ and $c_\mathrm{a}$ close to zero, which are in line with the portfolio loss correlation we determined in the context of Fig. \ref{fig:Copulas_Homogeneous_c=0_Differentns_cropped}. There, we discussed fluctuating asset correlations that occur for $N = 5$ and argued that the contributions of negative asset correlations to the portfolio loss are negligible. Hence, positive asset correlations dominate and cause the large portfolio loss correlations we just mentioned. For $\mu = -3\times10^{-3}\:\text{day}^{-1}$, however, the portfolio loss correlations depend only slightly on the ``fat-tail'' parameter $N$, as the impact of the strongly negative drift on the asset values and thus on the portfolio loss is much more important compared to the contribution of the involved random process.

\section{Simulation of Empirical Credit Portfolios}\label{sec:empirical}
After the previous parameter study of homogeneous portfolios, we now focus on more realistic portfolios with heterogeneous parameters. We will commence by studying the impact of choosing a single parameter heterogeneously and then turn to investigating realistic, fully heterogeneous portfolios. The section is concluded by an analysis of portfolio sizes and portfolio loss correlations.  
\subsection{Impact of Parameter Heterogeneity}\label{subsec:het}
\begin{figure}[h]
 \centering
 \includegraphics[width=0.75\textwidth]{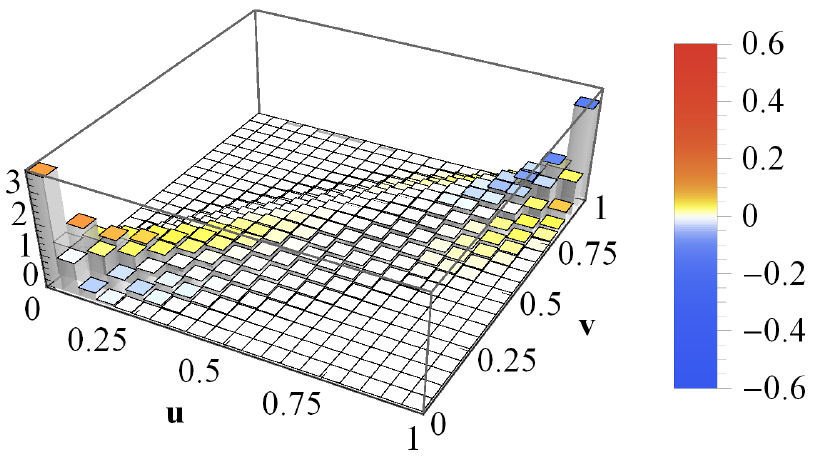}
  \caption{Averaged copula histogram of portfolio losses $L_1$ and $L_2$ of two portfolios with heterogeneous volatilities, $\sigma_i / \text{day}^{-1/2} \sim \mathcal{U}(0,0.25)$. The coloring indicates the local deviations from the Gaussian copula with $C_{L_1L_2}$.}
 \label{fig:Copula_Heterogeneous_Sigma_cropped}
\end{figure}
First, we consider a rather simple heterogeneous portfolio. It equals the homogeneous portfolio we have discussed so far, except for the fact that the daily volatilities $\sigma_i$ are not set to $\sigma_i = \sigma = 0.02\:\text{day}^{-1/2}$, but their values are drawn from the uniform distribution $\mathcal{U}(0,0.25)$. Moreover, it is important to mention that we choose the daily drift $\mu = -3\times 10^{-3}\:\text{day}^{-1}$ and the Gaussian limit ($N \rightarrow \infty$) of the ensemble averaged random process. The resulting portfolio loss copula for such portfolios is shown in Fig. \ref{fig:Copula_Heterogeneous_Sigma_cropped}. Due to the strongly negative drift, the probability of non-default events is zero. Nevertheless, we observe deviations from the Gaussian copula with parameter $C_{L_1L_2}$. This result is in contrast to the Gaussian copula displayed in the bottom panel of Fig. \ref{fig:Copulas_Homogeneous_DifferentMus_cropped}, where a homogeneous $\sigma = 0.02\:\text{day}^{-1/2}$ is assumed. For such a perfectly homogeneous portfolio, the Gaussian dependence of asset values is preserved at the level of portfolio losses. However, the Gaussian loss dependence is altered by the heterogeneous choice of one or more parameters; here, volatility. Thus, we have identified two causes of non-Gaussian copulas: parameter heterogeneity and the aforementioned projections of non-defaults onto zero.
\subsection{Simulation Setup and Results}\label{subsec:simresults}
\begin{figure}[H]
 \centering
 \includegraphics[width=0.65\textwidth]{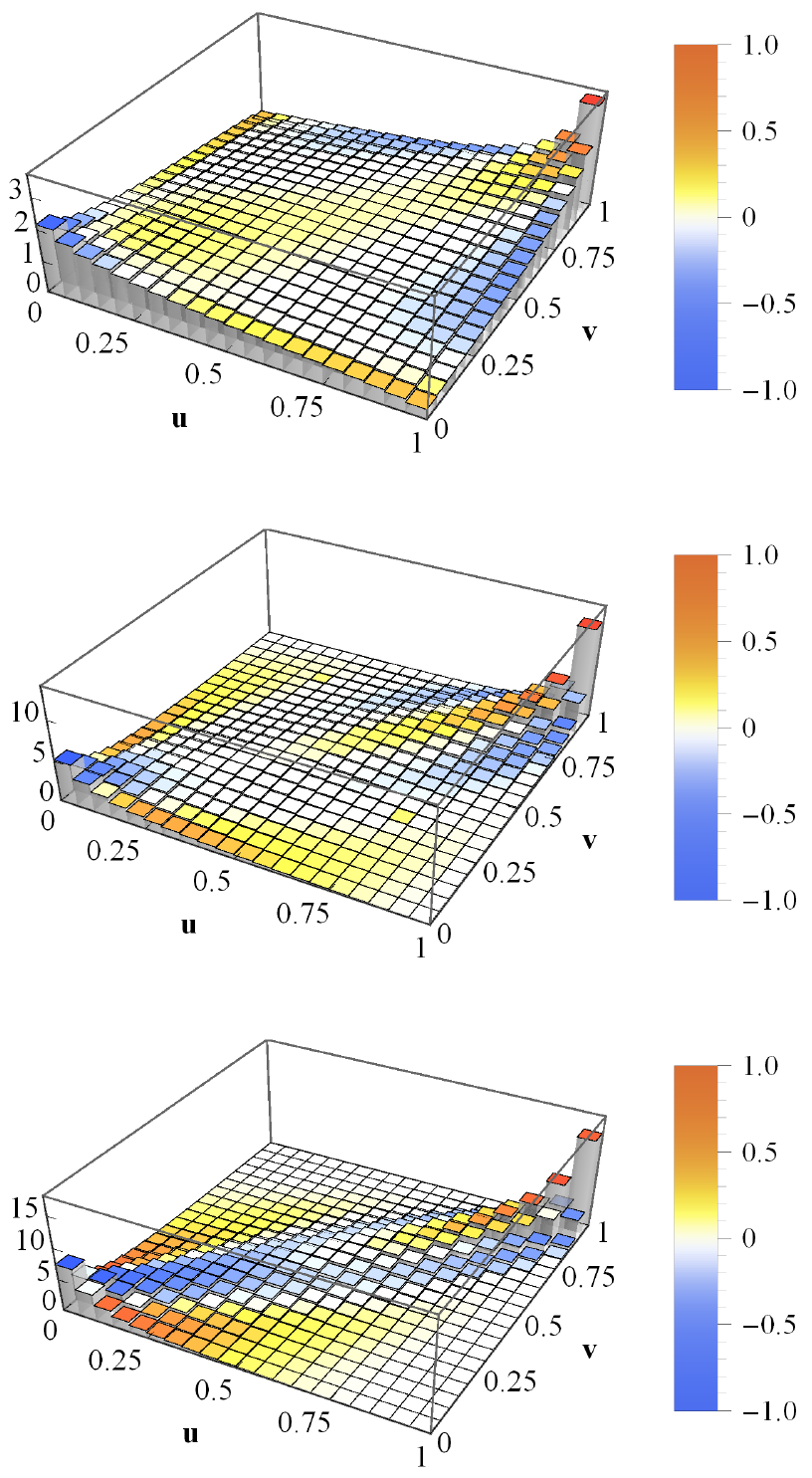}
  \caption{Time-averaged copula histogram of portfolio losses $L_1$ and $L_2$ of two empirical portfolios ($K = 50$). Top: portfolio 1 is always drawn from S\&P 500 and portfolio 2 from Nikkei 225, middle: both portfolios are drawn from S\&P 500, bottom: both portfolios are drawn from Nikkei 225. The coloring indicates the local deviations from the Gaussian copula with $C_{L_1L_2}$. The asset values are multivariate normal ($N \rightarrow \infty$).}
 \label{fig:Copula_LGD_Copula_Realistic_cropped}
\end{figure}

In order to set up realistic, fully heterogeneous portfolios we employ empirical data for drifts, volatilities and correlations. They are based on stock return data from S\&P 500 (272 companies) and Nikkei 225 (179 companies) and cover the period 01/1993--04/2014. We would like to obtain an average portfolio loss copula of empirical portfolios which is first averaged over many portfolio pairs and then averaged over the 21-year interval. To achieve this goal, we run the following procedure $20\,000$ times: First, an annual time interval within the 21-year period is randomly chosen. We determine the drifts, volatilities and correlations of all companies on this interval and draw two portfolios of size $K = 50$. Furthermore, the leverages $F_i/V_{0i}$ are drawn from the uniform distribution in  Eq.~(\ref{eq:levdist}). Finally, $10\,000$ portfolio loss simulations ($N \rightarrow \infty$) are run and the portfolio loss copula is estimated. This Monte-Carlo simulation enables us to determine the desired time-averaged portfolio loss copula of empirical portfolios. We consider three different cases: First, portfolio 1 is always drawn from S\&P 500, while portfolio 2 is always based on the Nikkei 225 (top panel of Fig. \ref{fig:Copula_LGD_Copula_Realistic_cropped}). Next, both portfolios are drawn from the S\&P 500 (middle panel), and finally, from Nikkei 225 (bottom panel). In the latter two cases we divided these stock markets into two ``sub-markets'' before drawing any portfolios. Thus, we avoid a given company to be part of portfolio 1 in one iteration and part of portfolio 2 in another iteration. Due to this division into ``sub-markets'', the almost perfect symmetry of all three copulas regarding the line spanned by $(0,0)$ and $(1,1)$ is not trivial, but an important feature. We will infer from Fig. \ref{fig:LossCorr_K_Realistic} that this behavior is due to the almost negligible idiosyncrasy for credit portfolios of size $K = 50$.

In the S\&P-Nikkei case, the deviations of the copula from the Gaussian copula are obvious, as we observe dependences of extreme events in the $(1,1)$-corner that are more pronounced than the Gaussian prediction. Further deviations are the narrower and more pointed $(1,1)$-tail as well as the flatter $(0,0)$-tail, which renders it quite asymmetric regarding the line spanned by $(0,1)$ and $(1,0)$. We interpret this result as follows. We find extreme
\goodbreak
\newpage
portfolio losses to occur more often simultaneously than it is the case for small portfolio losses. Thus, a modeling of portfolio loss dependences by means of Gaussian copulas is highly erroneous and might cause severe underestimations of the actual risks.

In the S\&P-S\&P as well as in the Nikkei-Nikkei case, the time-averaged copulas bear a certain resemblance to the result in the top panel. However, we find higher coupling strengths, as can be seen from Table \ref{tab:lossCorrs} which lists the average asset correlation and the portfolio loss correlation for all three cases.
\begin{table}
\centering
\begin{tabular}{|p{3cm}|p{2cm}|p{2cm}|}
  \hline
  \multicolumn{1}{|c|}{Stock Markets} & \multicolumn{1}{|c|}{$c_\mathrm{a}$} & \multicolumn{1}{|c|}{$C_{L_1L_2}$}\\
  \hline
  \hline
  S\&P-Nikkei & $0.119$ & $0.287$ \\
  \hline
  S\&P-S\&P & $0.266$ & $0.779$ \\
  \hline
  Nikkei-Nikkei & $0.385$ & $0.927$ \\
  \hline
\end{tabular}
  \caption{Averaged asset correlation $c_\mathrm{a}$ and portfolio loss correlation $C_{L_1L_2}$ of empirical portfolios drawn from the indicated stock markets.}
  \label{tab:lossCorrs}
\end{table}
\begin{figure}[H]
 \centering
 \includegraphics[width=0.75\textwidth]{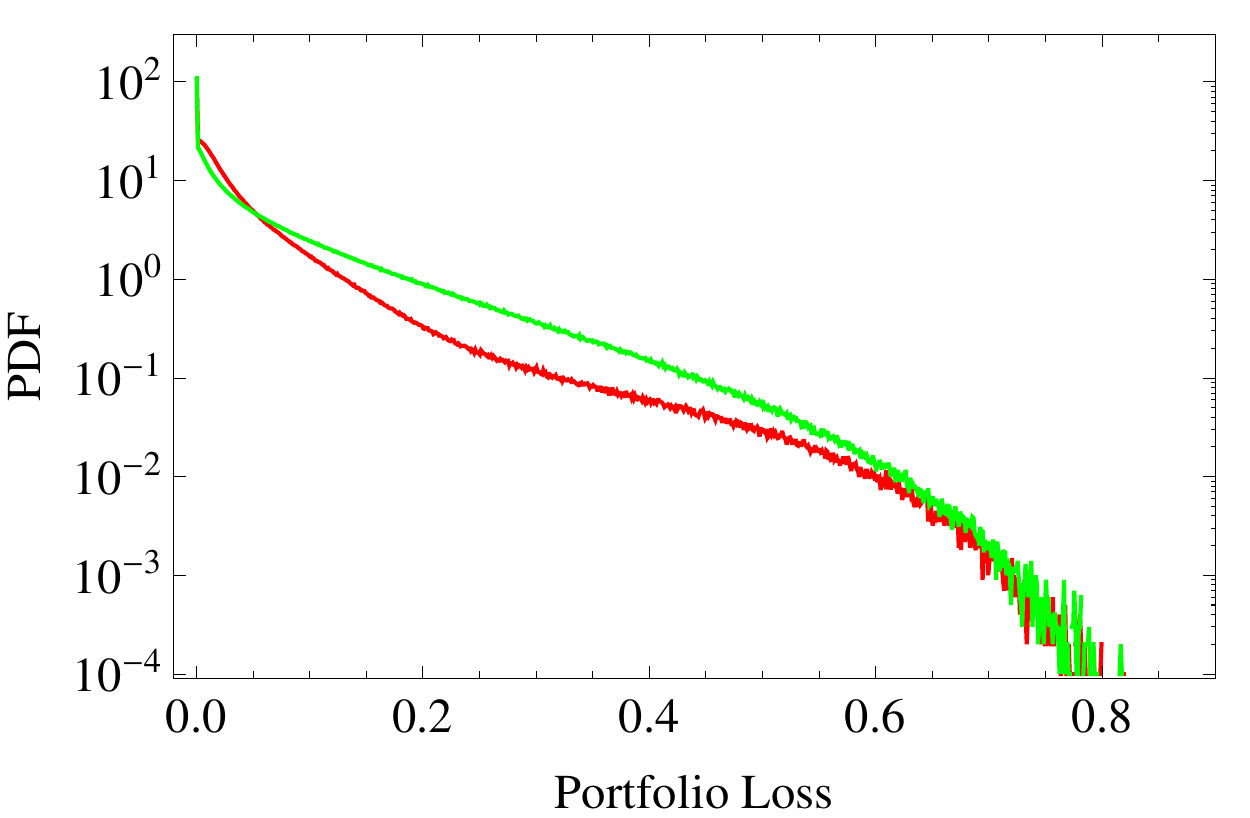}
  \caption{Semi-log scaled time-averaged portfolio loss pdfs of empirical portfolios of size $K = 50$. The portfolios are drawn from S\&P 500 (red) and Nikkei 225 (green), respectively. The asset values are multivariate normal ($N \rightarrow \infty$).}
 \label{fig:Averaged_LossDistribution_Empirical_K=50}
\end{figure}
For $K = 50$, the averaged marginal distributions of empirical portfolio losses are depicted in Fig. \ref{fig:Averaged_LossDistribution_Empirical_K=50}. The S\&P 500-related pdf is shown in red, the Nikkei 225-related result in green. They are highly asymmetric and exhibit distinct tails as well as delta peaks at zero. These features are in accordance with the blue and red curve in Fig. \ref{fig:LossDistributions_Homogeneous_DifferentMus}, which illustrate the portfolio loss distributions of a homogeneous portfolio with drifts  $\mu = 10^{-3}\:\text{day}^{-1}$ (blue) and $\mu = 3\times10^{-4}\:\text{day}^{-1}$ (red), respectively. In contrast, the averaged empirical portfolio loss pdfs have much heavier tails. While the risk of extreme portfolio losses ($L > 0.7$) is equally high for S\&P- and Nikkei-based portfolios, the Nikkei-based portfolio loss pdf is almost half a magnitude higher for large portfolio losses ($0.15 < L < 0.45$). Considering the delta peak more closely, we find the time-averaged probabilities of portfolio non-default events to be $11.7\%$ (S\&P 500) and $12.2\%$ (Nikkei 225), respectively. 

\subsection{Portfolio Size and Portfolio Loss Correlation}\label{subsec:portsize}
Although we have seen that a Gaussian modeling of loss dependences falls short in most cases, it can be seen as a rough but useful approximation of the actual dependence. Especially, as such a reduction of the full statistical dependence to a single number simplifies the study of further parameter dependences. Here, we consider the average loss correlation for empirical portfolios as a function of portfolio size $K$. So far, we have discussed $K = 50$ and have found portfolio loss correlations which were much smaller than the ones for homogeneous portfolios of the same size. This is unsurprising, because empirical portfolios are heterogeneous and thus exhibit higher idiosyncrasies for a given portfolio size.
\begin{figure}[h]
 \centering
 \includegraphics[width=0.73\textwidth]{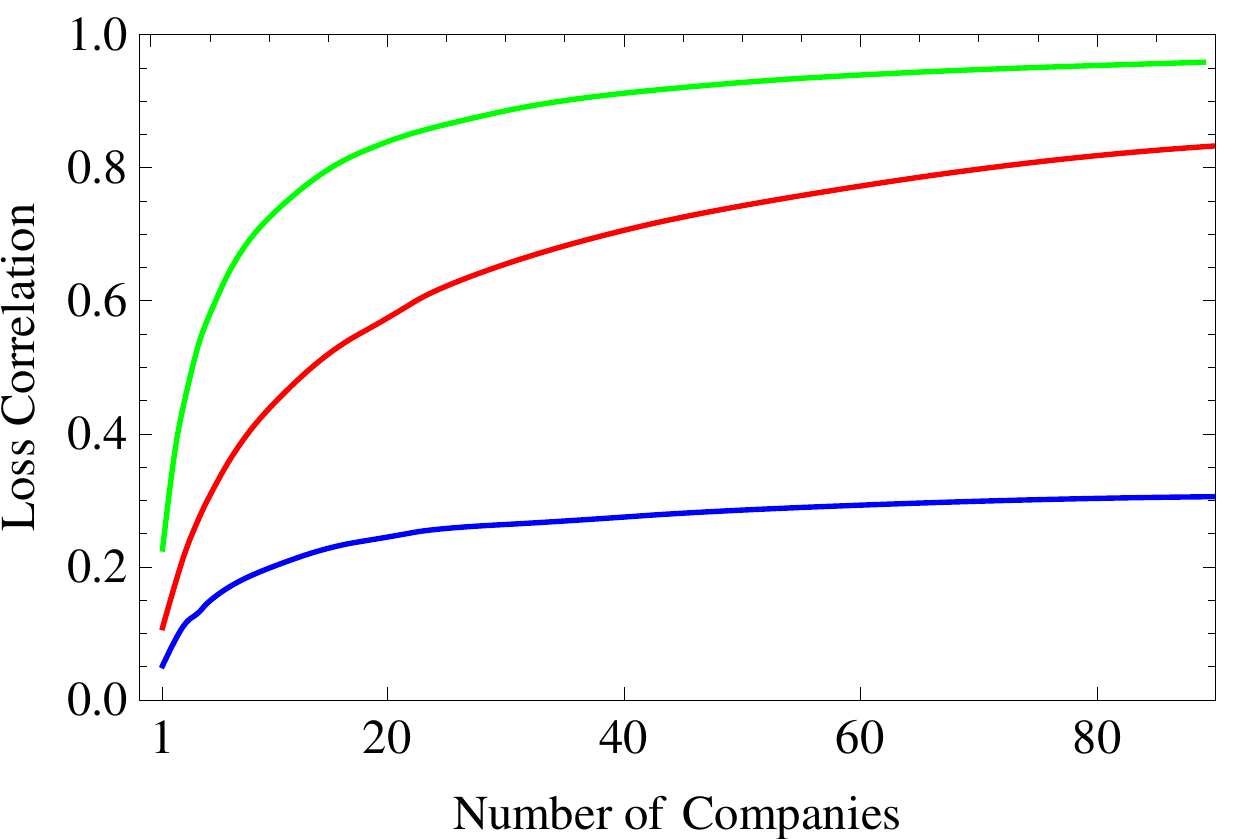}
  \caption{Time-averaged portfolio loss correlation as a function of portfolio size $K$ for empirical portfolios. Red: both portfolios are drawn from S\&P 500, green: both portfolios are drawn from Nikkei 225, blue: portfolio 1 is always drawn from S\&P 500 and portfolio 2 always from Nikkei 225.}
 \label{fig:LossCorr_K_Realistic}
\end{figure}
Once again, we distinguish three cases: Both portfolios are S\&P 500-based, both are Nikkei 225-based, one is S\&P 500-based and the other one Nikkei 225-based. Taking a look at Fig. \ref{fig:LossCorr_K_Realistic}, we observe steep increases in portfolio loss correlation with increasing $K$, which is in line with our findings for homogeneous portfolios. This behavior is due to the decreased idiosyncrasy of large portfolios, which causes very similar portfolio losses and, thus, high portfolio loss correlations. In all three cases, idiosyncrasy is almost negligible for $K \geq 80$.

The results in Fig. \ref{fig:LossCorr_K_Realistic} are of high practical relevance, as the underlying portfolios are based on realistic, empirically determined parameter sets. For S\&P 500-based portfolios we find, in particular, loss correlations $C_{L_1L_2} > 0.5$ for $K \geq 14$, $C_{L_1L_2} > 0.7$ for $K \geq 40$ and $C_{L_1L_2} > 0.85$ for $K \geq 150$ (not shown here). This means that losses of medium-sized disjoint portfolios ($K = 150$) are almost perfectly correlated and that even small ($K = 40$) and very small ($K = 14$) portfolios exhibit rather strong coupling. Thus, high dependences among banks are not limited to ``big players'', which hold portfolios of several thousand credit contracts, but affect small financial institutions in a very similar fashion.

In the Japanese stock market we observe even higher portfolio loss correlations, as the underlying asset correlations are on average higher compared to the US market (see above). The opposite is the case if one Japanese and one US-American portfolio is considered. Here, we find considerably weaker asset correlations and thus loss correlation coefficients $C_{L_1L_2} \leq 0.35$.

\section{Discussion}\label{sec:discussion}

We addressed the dependence of concurrent credit portfolio losses using Monte-Carlo simulations within the framework of the Merton model.
For two non-overlapping credit portfolios, we estimated the copulas of portfolio losses to reveal their full dependence structure.  
We found concurrent large portfolio losses to be more likely than concurrent small portfolio losses. These deviations from an ideally Gaussian behavior could be traced back to two different causes: the dominance of non-default events and the heterogeneity of empirical portfolios. Thus, concurrent severe losses of credit portfolios are underestimated if employing only standard correlation coefficients. In contrast, copulas allow for a more realistic view of such extreme events and the systemic risk they pose. 
Looking at portfolio loss correlations from another perspective, we found this wide-spread measure to exhibit further weaknesses, as we exposed its dependence on the ratio of default and non-default events. If constant asset correlations between single credits or credit portfolios are given, the correlations of losses are higher the worse the market period.
Finally, we analyzed how portfolio loss dependences scale with portfolio size and found surprisingly strong couplings of portfolio losses even for medium-sized and small disjoint empirical portfolios.
In summary, we took a new inter-portfolio perspective of credit risk and thereby provided further insight into intrinsic instabilities of the financial sector. We believe that the interdependences of tail risks are of fundamental importance for the fixed-income market and thus relevant for both regulators and investors.    

\section*{Declaration of Interest}
The authors report no conflict of interest. The authors alone are responsible for the content and writing of the paper.


\end{document}